\documentclass[preprintnumbers, floatfix, preprintnumbers, letterpaper, twocolumn, superscriptaddress,nofootinbib]{revtex4}
\usepackage{graphicx}
\usepackage{microtype}
\usepackage{amsmath}
\usepackage{amssymb}
\usepackage{subfigure}
\usepackage{hyperref}
\usepackage{url}
\usepackage{xcolor}
\usepackage{color}
\usepackage{mathrsfs}
\usepackage{calrsfs}
\usepackage{amsfonts}
\usepackage{tabularx}
\usepackage{eucal}
\usepackage{latexsym}
\usepackage{ragged2e}
\usepackage{epsfig}
\usepackage{textcomp}
\usepackage{float}

\usepackage{caption}
\DeclareCaptionJustification{justified}{\leftskip=0pt \rightskip=0pt \parfillskip=0pt plus 1fil}
\definecolor{vividviolet}{rgb}{0.62, 0.0, 1.0}
\definecolor{amaranth}{rgb}{0.9, 0.17, 0.31}
\definecolor{palatinateblue}{rgb}{0.15, 0.23, 0.89}
\definecolor{brightpink}{rgb}{1.0, 0.0, 0.5}
\definecolor{cornflowerblue}{rgb}{0.39, 0.58, 0.93}
\definecolor{deepcarminepink}{rgb}{0.94, 0.19, 0.22}
\definecolor{radicalred}{rgb}{1.0, 0.21, 0.37}

\hypersetup{ linktoc=all,
	colorlinks, linkcolor={palatinateblue},
	citecolor={brightpink}, urlcolor={amaranth}
}

\graphicspath{{Images/}}

\def\sideremark#1{\ifvmode\leavevmode\fi\vadjust{\vbox to0pt{\vss
			\hbox to 0pt{\hskip\hsize\hskip1em
				\vbox{\hsize1.3cm\tiny\raggedright\pretolerance10000
					\noindent #1\hfill}\hss}\vbox to8pt{\vfil}\vss}}}%
\def\beq{\begin{equation}}
\def\eeq{\end{equation}}

\begin{document}

\title{The Hookean Law of Black Holes and Fragmentation:\\ Insights from Maximum Force Conjecture and Ruppeiner Geometry}

\author{Sofia Di Gennaro}
\email{sofia.digennarox@gmail.com}
	\affiliation{Center for Gravitation and Cosmology, College of Physical Science and Technology, Yangzhou University, \\180 Siwangting Road, Yangzhou City, Jiangsu Province  225002, China}
	
\author{Michael R. R. \surname{Good}}
\email{michael.good@nu.edu.kz}
\affiliation{Department of Physics \& Energetic Cosmos Laboratory, Nazarbayev University, Nur-Sultan 010000, Kazakhstan}

\author{Yen Chin \surname{Ong}}
\email{ycong@yzu.edu.cn}
\affiliation{Center for Gravitation and Cosmology, College of Physical Science and Technology, Yangzhou University, \\180 Siwangting Road, Yangzhou City, Jiangsu Province  225002, China}
\affiliation{Shanghai Frontier Science Center for Gravitational Wave Detection, School of Aeronautics and Astronautics, Shanghai Jiao Tong University, Shanghai 200240, China}

\begin{abstract}
We show that the notion of ``Hookean law'' $F=kx$, suitably defined in asymptotically flat singly spinning Myers-Perry black hole spacetimes in dimensions $d\geqslant 5$, is related to the Emparan-Myers fragmentation (splitting of a black hole into two becomes thermodynamically preferable). Specifically, the values of black hole parameters when fragmentation occurs correspond to the maximal value of $F$. Furthermore this always happens before $F$ reaches $1/4$ in Planck units. These results suggest that a version of ``maximum force conjecture'' may be relevant for black hole thermodynamics. We also relate these findings to the Ruppeiner thermodynamic geometry of these black holes and speculate on the implications for the underlying microstructures of black hole horizons.  
\end{abstract} 

\maketitle

\section{Introduction: The Maximum Force Conjecture and Black Hole Thermodynamics}

The maximum force conjecture proposed by Gibbons \cite{0210109, 1408.1820} and Schiller \cite{0607090, 724159} argues that in general relativity any physically attainable force or tension between two bodies is bounded above by a quarter of the Planck force $F_{\text{pl}}$. In 4-dimensions, the value in SI units is
\begin{equation}
F \leqslant \frac{c^4}{4G} = \frac{1}{4}F_{\text{pl}} \approx 3.25 \times 10^{43} ~\text{N}.
\end{equation}
Note that this is a classical quantity free of $\hbar$. 
Hereinafter we will work in Planck units in which $\hbar=G=c=k_B=1$ for simplicity of notation (so $F_{\text{pl}}=1$); here $k_B$ denotes the Boltzmann constant. 

The inequality above is the strong form of the conjecture, whilst the weak form allows the maximum force to be some $\mathcal{O}(1)$ multiple of the Planck force, not necessarily $1/4$ \cite{1809.00442}. The conjecture in either form has been shown to be incorrect by Jowsey and Visser \cite{2102.01831}, who provided a few counterexamples (such as fluid spheres on the verge of gravitational collapse). Nevertheless, as mentioned therein, the conjecture could be correct under certain specific conditions. That is, perhaps the discussion must be restricted to some classes of forces. Subsequently Faraoni \cite{2105.07929} proposed that the idea of maximum force does apply in the context of black holes, but argued that such a bound on any force acting on black hole horizons is inherently tied to cosmic censorship (and therefore not a new, independent conjecture). Schiller has recently objected to these counterexamples and arguments \cite{2109.07700, 2112.15418}, and further clarified the need to distinguish between local and non-local forces (see also the follow-up of Faraoni \cite{PhysRevD.104.068502}). It is clear that the current state of the conjecture -- not just its validity but also its precise formulation and the contexts in which it is of relevance -- remains debatable, and thus further explorations are necessary. See also \cite{2005.06809,2006.07338}.

In \cite{1412.5432} a ``Hookean force'' of the form $F=kx$ was introduced in the context of 4-dimensional asymptotically flat Kerr black holes, in which $k$ is a ``spring constant'' defined by $k=M\Omega_+^2$ (we will review this in Sec.(\ref{2})), while $x$ is naively identified with the horizon $r_+$. 
\emph{We caution the readers that $F$ does not have the right physical dimension for a force.} 
Indeed $F$ is always classical -- as will become clear later, whereas a force in $d$-dimension contains the Planck constant $\hbar$; surprisingly a force is necessarily quantum in higher dimensions (explicitly, the Planck force is \cite{2005.06809} $F_{\text{pl}} = G^{\frac{2}{2-d}}c^{\frac{4+d}{d-2}}\hbar^{\frac{4-d}{2-d}}$).

Therefore, strictly speaking we are not directly relating the Hookean law to the maximum force conjecture. However, by multiplying a force
with the appropriate powers of $\hbar, c$ and $G$ we can always construct the relevant physical quantities, and thus extending the maximum force conjecture to these quantities instead\footnote{Although we should not expect the $1/4$ coefficient to always carry over, usually an $\mathcal{O}(1)$ factor is expected.}. 
For example, the maximum luminosity conjecture \cite{2112.15418, 724159, 1803.03271, 2105.06650} was obtained by multiplying a factor of the speed of light $c$ to the force ($L \lesssim c^5/G \approx 3.6 \times 10^{52}~\text{W}$, the so-called\footnote{See \cite{2105.06650} for the origin of this mis-attribution.} ``Dyson luminosity''), as well as the related maximum mass loss rate $\sim c^3/G$ \cite{2109.05973}. 
Indeed, instead of a collection of maximum conjectures that apply to various physical quantities, we feel that it is most convenient to just refer to the maximum force conjecture, with the implicit understanding that the appropriate powers of $\hbar, c$ and $G$ should be applied for conversion. While working in Planck units, all quantities are of course dimensionless, but these caveats should be kept in mind. We will henceforth drop the scare quote around the ``Hookean force''.

It was shown in \cite{1412.5432} that the Hookean force tends to $1/4$ in the extremal limit (since $\Omega_+ = a/(r_+^2+a^2) \leqslant 1/2M$):
\begin{equation}
\lim_{J\to M^2} F = \frac{1}{4M}\left(\lim_{J\to M^2}  r_+\right)=\frac{1}{4M} \cdot M =\frac{1}{4},
\end{equation}
which seems to suggest that the maximum force conjecture is indeed related to cosmic censorship, as argued by Faraoni in \cite{2105.07929} (though see his different view in \cite{PhysRevD.104.068502}). 
Indeed $F$ increases from 0 to $1/4$ as the angular momentum increases from zero to its extremal value.
However, in general the maximum force conjecture in higher dimensions need not coincide with the censorship bound, as we shall see below.
Incidentally, let us mention that the curvature singularity of the extremal Kerr black hole in the massless limit can be interpreted as an infinitely thin cosmic string loop with negative tension $T=-1/4$, whose absolute value satisfies the maximum force conjecture \cite{1705.07787,1606.04879}.

When generalized to spacetime dimension $d$ above 4 the picture is quite different. Let us note that applying the maximum force conjecture to higher dimensions is actually quite subtle due to the result in \cite{2005.06809}, an issue we will defer to the Discussion section. At this point, we shall emphasize that our objective is \emph{not} to prove the conjecture in any form, but rather to argue for a surprising relationship between the Hookean law and thermodynamic instability, while also connect this with the conjectured $1/4$ value.

In higher dimensions, there can be more than one angular momenta, but we will focus on the singly rotating case in which there is \emph{no} extremal limit in $d \geqslant 6$, though much of the discussion applies also to the 5-dimensional case. We find that the Hookean force is bounded by a dimensional dependent value less than $1/4$, which corresponds to the well-known Emparan-Myers fragmentation \cite{0308056,0907.2248} -- black holes that are spinning too fast break into two because the latter configuration has higher entropy and is therefore thermodynamically preferred\footnote{In 4 dimensions black holes \emph{cannot} break apart. This is a consequence of Hawking area theorem \cite{hawking}, but the underlying physical reason is the same: configuration of two black holes in this case would have a lower Bekenstein-Hawking entropy.}. Thus, remarkably the Hookean force detects thermodynamic instability: in some sense the black hole breaks much like a spring would if stretched too much.

\section{Springy Black Holes and the Hookean Force}\label{2}

It was discovered in \cite{1412.5432} that the Hawking temperature of an asymptotically flat Kerr black hole in 4 dimensions can be expressed succinctly as
\begin{equation}\label{temp}
T = \frac{1}{2\pi}(g - k),
\end{equation}
in which $g:=(4M)^{-1}$ is the surface gravity of a Schwarzschild black hole of the same mass.  The ``spring constant'' $k$ is defined by $k:=M\Omega_+^2$, in analogy with the familiar expression $k=m\omega^2$ for a mass $m$ attached to the end of a spring in elementary classical mechanics\footnote{To keep the analogy we prefer to keep (the hidden) $\hbar$ outside the bracket in Eq.(\ref{temp}), so that $k$ -- and consequently the Hookean force -- is a classical quantity. 
Explicitly, in conventional units, $T=\frac{\hbar}{2\pi ck_B}\left[\frac{c^4}{4GM}-\frac{G}{c^2}M\Omega_+^2\right].$
We will also do this in higher dimensions.}. 
Let us now consider higher dimensional rotating black holes (the Myers-Perry solution \cite{MP}) with one angular momentum and use Eq.(\ref{temp}) to define $k$.

The Hawking temperature of a Myers-Perry black hole in $d$-dimensional spacetime is given by
\begin{equation}
	T = \frac{1}{4\pi}\left(\frac{2r_+^{d-4}}{\mu}+ \frac{d-5}{r_+}\right), \label{mp temp}
\end{equation}
where $\mu$ is essentially a normalized mass parameter\footnote{Note that $\Omega_{d-2}$ is the area of the unit $(d-2)$-dimensional sphere, not to be confused with the angular velocity of the horizon $\Omega_+$.
}:
\begin{equation*}
	\mu := \frac{16\pi G}{(d-2)\Omega_{d-2}} M, \qquad \Omega_{d-2} := \frac{2\pi^{\frac{d-1}{2}}}{\Gamma\left(\frac{d-1}{2}\right)}.
\end{equation*}

The horizon is the solution of the expression
\begin{equation}
	r_+^2 + a^2 - \frac{\mu}{r_+^{d-5}} = 0. \label{mp hor}
\end{equation}
In $d=5$, the extremal limit corresponds to $r_+=0$; it is an example of the so-called ``extremal vanishing horizon'' (EVH) spacetime \cite{1107.5705}. Since the horizon overlaps with the central ring singularity, the extremal solution is essentially a naked singularity \cite{0308056}. In $d \geqslant 6$ dimensions, singly rotating Myers-Perry black holes can possess arbitrarily large angular momentum since there is no extremal limit\footnote{Note that in $d\geqslant 5$ singly-spinning Myers-Perry black holes have only one horizon. They can only become extremal in the EVH sense, which only occurs in 5 dimensions.}.
The case $a\gg r_+$ is called the ``ultra-spinning limit''. As such we expect that the case for $d \geqslant 6$ would be substantially different from the $d=5$ and $d=4$ cases.

In the ultra-spinning limit, we have, from Eq.(\ref{mp hor}),
\begin{equation}
	\mu = r_+^{d-5}a^2. \label{mp hor us}
\end{equation}
We can substitute this into Eq.\eqref{mp temp} and obtain:
\begin{equation}
	T = \frac{1}{2\pi}\left(\frac{r_+}{a^2}+ \frac{d-5}{2r_+}\right). \label{mp temp us}
\end{equation}
The requirement that $T>0$ is guaranteed even in the ultra-spinning limit $a \gg r_+$ if $d>5$. 
Next we consider the temperature of a higher dimensional Schwarzschild-Tangherlini black hole \cite{tang} and denote its horizon by $r_{+s}$. To do so, we take the limit $a\to 0 $ in Eq.\eqref{mp hor} and we obtain $\mu = r_{+s}^{d-3}$. Then we substitute it into Eq.\eqref{mp temp}:
\begin{equation}
	T_s = \frac{d-3}{4\pi r_{+s}} = \frac{d-3}{4\pi \mu^{\frac{1}{d-3}}}. \label{mp temp sch}
\end{equation}
Following \cite{1412.5432} and using the fact that the surface gravity is $\kappa  = 2\pi T$, we want to rewrite Eq.\eqref{mp temp us} in the form of Eq.(\ref{temp}). Using Eq.\eqref{mp hor us}, \eqref{mp temp sch} and \eqref{mp temp us}, we have:
\begin{align}
	T & = \frac{1}{2\pi}\left[\frac{d-3}{2 \mu^{\frac{1}{d-3}}} - \left(\frac{d-3}{2\mu^{\frac{1}{d-3}}} -\frac{d-5}{2r_+} - \frac{r_+}{a^2}\right)\right] \nonumber\\
	& = \frac{1}{2\pi}\left\{\frac{d-3}{2 \mu^{\frac{1}{d-3}}} - \underbrace{\left[
	\frac{d-3}{2r_+} \left(\frac{r_+}{a}\right)^{\frac{2}{d-3}}-\frac{d-5}{2r_+} - \frac{r_+}{a^2} \right]}_{=:k}\right\},\label{temp mp gen}
\end{align}
where we identify the quantity in the square brackets of the second line as the spring constant.
The Hookean force $F := kx$ with $x = r_+$, then yields
\begin{equation}
	F = \frac{d-3}{2} \left(\frac{r_+}{a}\right)^{\frac{2}{d-3}}-\frac{d-5}{2} - \frac{r_+^2}{a^2} \lesssim 0.21, \label{force us}
\end{equation}
in which the upper bound value was obtained numerically given that $r_+/a < 1$ and $d>5$ are satisfied. These inequalities are the conditions required by the ultra-spinning limit and the positiveness of the Hawking temperature. {(The results do not change even if we are working with the full expressions without taking the ultra-spinning limit.)

We observe that the upper bound 0.21 is less than the conjectured 1/4. What is more surprising is that this value also corresponds to the Emparan-Myers fragmentation, as can be checked explicitly. 
Leaving $\mu$ explicit, the Hawking temperature is
\begin{equation*}
	T = \frac{1}{2\pi}\left[\frac{d-3}{2\mu^{\frac{1}{d-3}}} - \left(\frac{d-3}{2\mu^{\frac{1}{d-3}}} - \frac{d-5}{2r_+}-\frac{r_+^{d-4}}{\mu}\right)\right].
\end{equation*}
The Hookean force is then expressible as
\begin{equation}\label{hook}
	F = k r_+ = \frac{d-3}{2}\frac{r_+}{\mu^{\frac{1}{d-3}}} -\frac{r_+^{d-3}}{\mu}  - \frac{d-5}{2}.
\end{equation}
In 6 dimensions, the approximate result in \cite{0308056} is that the black hole decays into two Schwarzschild fragments when $a/r_+ > 1.36$.
Thus requiring that
\begin{equation}
	\frac{a}{r_+} \lesssim 1.36  \qquad \implies \qquad \frac{a^3}{\mu} \lesssim 0.88,\label{cond d6}
\end{equation}
and substituting this into the equation for the force yields:
\begin{equation}
	F  = \frac{3}{2}\frac{r_h}{\mu^{\frac{1}{3}}} -\frac{r_+^{3}}{\mu} -\frac{1}{2} = \frac{3}{2}\frac{r_+}{a}\frac{a}{\mu^{\frac{1}{3}}} -\frac{r_+^{3}}{a^3}\frac{a^3}{\mu} -\frac{1}{2} \lesssim 0.21. \notag 
\end{equation}
This agrees with Eq.(\ref{force us}) despite ${a}/{r_+} \leqslant 1.36$ is arguably not too ultra-spinning.
Similarly, in $7$ and $8$ dimensions we have $F \lesssim 0.19$ and $F \lesssim 0.18$, respectively, which are less than the 0.21 value obtained in the ultra-spinning limit in Eq.\eqref{force us}.
It can be shown that, in fact, the upper bound decreases monotonically with $d$, and tends to 0.1534 in the limit $d \to \infty$.

In 5-dimensions, we can solve Eq.(\ref{mp hor}) \emph{exactly} and obtain the expression for the horizon
\begin{equation}
	r_+ = \sqrt{\mu - a^2}.  \label{hor 5}
\end{equation}
The temperature in Eq.(\ref{mp temp}) and the temperature in the Schwarzschild limit in Eq.\eqref{mp temp sch} become, respectively,
\begin{equation*}
	T = \frac{r_+}{2\pi\mu}, \qquad T_s = \frac{1}{2\pi r_{+s}} = \frac{1}{2\pi\sqrt{\mu}}. 
\end{equation*}
The temperature is always positive, provided the cosmic censorship bound $\mu \geqslant a^2$ holds. Using the expression for the horizon in Eq.\eqref{hor 5}, we obtain:
\begin{align*}
	T & = \frac{1}{2\pi} \left[ \frac{1}{\sqrt{\mu}} - \left(\frac{\sqrt{\mu}-\sqrt{\mu - a^2}}{\mu}\right)\right] = \frac{1}{2\pi} \left(g -k \right).
\end{align*}
Therefore, the Hookean force is given by:
\begin{flalign}
	F &= kr_h = \left(\frac{\sqrt{\mu}-\sqrt{\mu - a^2}}{\mu}\right)  \sqrt{\mu - a^2} \notag \\ 
	&= \sqrt{1-\frac{a^2}{\mu}}\left(1-\sqrt{1-\frac{a^2}{\mu}}\right) \leqslant \frac{1}{4},
\end{flalign}
where the last inequality is obtained by maximizing the function $F$ with respect to $a^2/\mu$. This again agrees with the fragmentation values approximated in \cite{0308056}, namely $a/r_+ \approx1.6$ and $a^2/\mu\approx0.72$, via Eq.(\ref{hook}).

We can therefore conclude that: in dimensions 4 and 5, in which there exists an extremal limit for singly spinning black holes, the maximum Hookean force is $1/4$. However this maximum value does \emph{not} correspond to the extremal limit in dimension 5 (as it does in dimension 4), since in that case $\mu=a^2$ and $F=0$. This makes sense because $r_+=0$ for the extremal 5-dimensional solution.
The maximum value of the Hookean force is attained at $a^2/\mu = 3/4$ instead. This serves as one counter-example that maximum force corresponds to cosmic censorship.
For dimension 6 and above there is no extremal limit, and the exact bound on the Hookean force is dimensional dependent and decreasing with number of dimensions. 
Furthermore the bound corresponds to the Emparan-Myers fragmentation point in each given dimension.
{\it{Thus the Hookean force reveals thermodynamical instability in black holes.}}  This is despite its seemingly naive definition.

One would naturally ask: does fragmentation always happen before $F=kx$ attains $1/4$? The answer is no in general (this non-universality is not necessarily a shortcoming -- see Discussion). For example, fragmentation of black holes with a Gauss-Bonnet term was studied in \cite{1412.4189}. Although the black hole entropy receives a correction, its temperature remains unchanged from Schwarzschild, so we cannot define $k$ in this case. It may be that this correspondence only holds in asymptotically flat spacetimes in general relativity without matter or with specific form of matter (as the Hookean force for the charged case is not as readily defined \cite{1412.5432}). 

\begin{figure}[h!] 
\centering
\includegraphics[width=.42\textwidth]{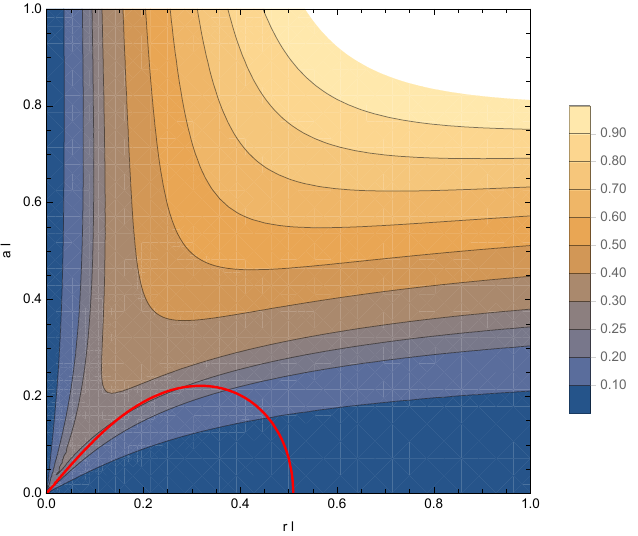}
\caption{The contour plot of the Hookean force $F=kx$ for AdS-Kerr black hole in 5-dimensions and the plot of the fragmentation curve (in red) along which the black hole splits into two asymptotically flat Schwarzschild black holes. The fragmentation always occurs before the maximum value $F=0.2570 \pm 0.0001 \approx 1/4$ is attained. Here the horizontal axis is $r_+l$ where $l$ is the inverse of the AdS radius, while the vertical axis is $al$. \label{ads}}
\end{figure}

In the presence of a negative cosmological constant, a 5-dimensional anti-de Sitter (AdS-)Kerr black hole can fragment into two AdS-Schwarzschild black holes, but for some values of the parameters, it can be checked that fragmentation occurs at $F>1/4$. However, if the same AdS-Kerr black hole were to fragment into two \emph{asymptoticaly flat} Schwarzschild black holes, then the maximum Hookean force at which fragmentation happens \emph{does} equal\footnote{This is not exact, but come extremely close: $F=0.2570 \pm 0.0001$. In any case the fragmentation curve itself is not exact (see \cite{0308056}) so it is possible that the maximum Hookean force value is $F\lesssim 1/4$.} $1/4$, see Fig.(\ref{ads}). The fact that $F=1/4$ appears in such a non-trivial example is a strong evidence that there is a deeper relation between fragmentation and the maximum force conjecture albeit not in a univesal fashion. In $d\geqslant 6$, the value of the maximum force decreases (and thus, is less than $1/4$).
In this example the fragmentation process involves a varying cosmological constant, unusual but may be possible if one considers the extended black hole thermodynamics, in which the cosmological constant is treated as (the negative of) a pressure term in the first law of black hole thermodynamics \cite{1209.1272, 0904.2765, 1404.2126, 1608.06147}, which can be varied\footnote{A recent work by Dai et al. gives a concrete mechanism that allows such a change -- via bubble nucleation \cite{2111.03359}. However, negative energy density is required for the bubble wall if the value of the cosmological constant were to increase. Nevertheless, it is conceivable that other processes might exist that permit such a transition.}.

\section{Thermodynamic Geometry and the Maximum Force}

\begin{figure*}[!t]
\begin{tabular}{rrr}
    \includegraphics[width=.33\textwidth]
      {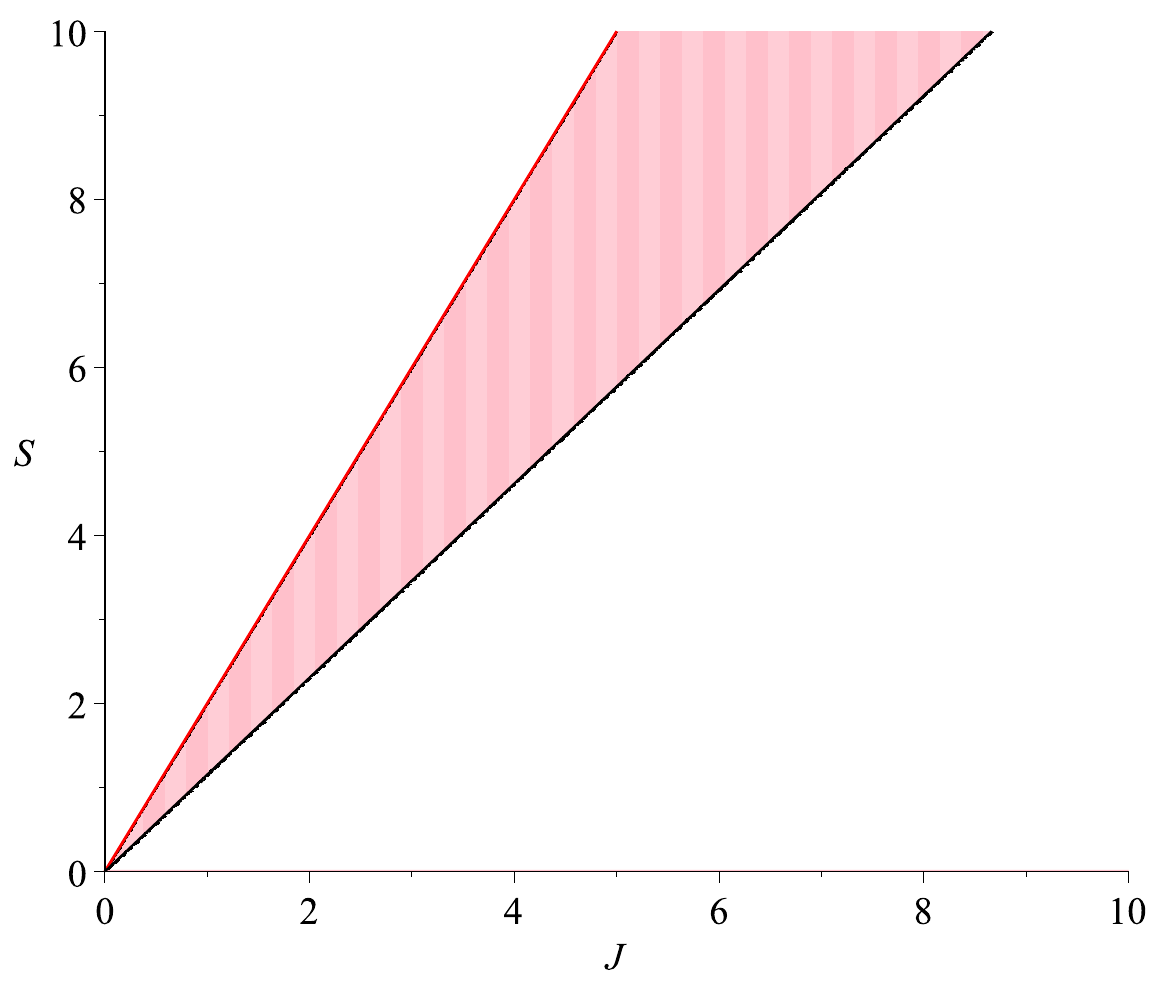}
    \includegraphics[width=.33\textwidth]
      {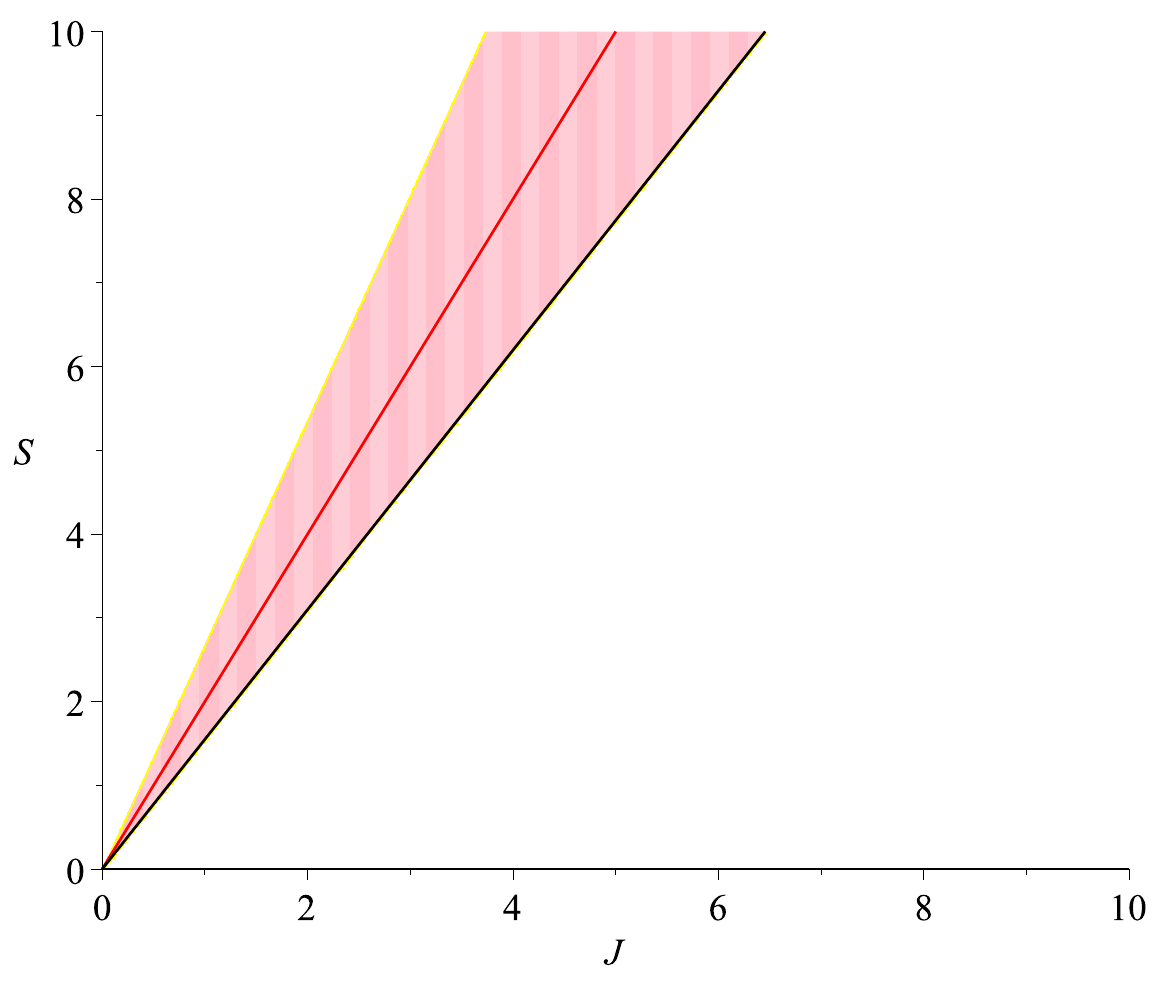}
		\includegraphics[width=.33\textwidth]
		  {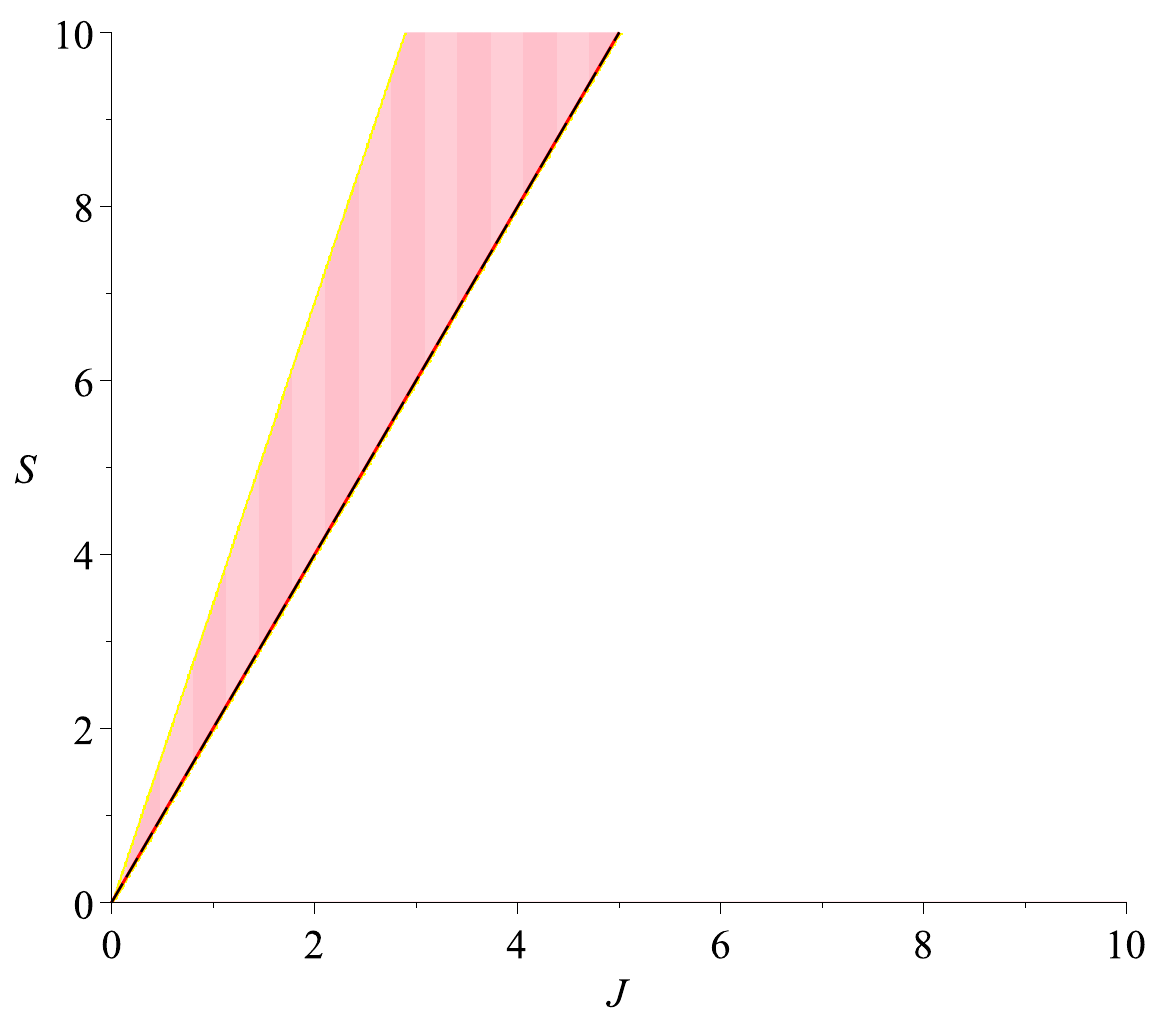}
\end{tabular}
	\caption{The shaded regions in these plots have positive Ruppeiner scalar curvature. Its boundaries are marked by yellow lines. From left to right we have the plots for $d=6$, $d=8$, and $d=10^4$, respectively. The red line corresponds to $F_2=1/4$. The black line corresponds to $F_2=\frac{1}{4} \left(\frac{d-3}{d-5}\right)$ along which $\mathcal{R}$ is divergent (phase transition). The black line always marks the lower boundary  of the positive $\mathcal{R}$ region. (By upper and lower boundaries we are referring to the positions in the plots, not the numerical values of $F_2$. In fact the lower boundary has a higher value of $F_2$ compared to the upper boundary.) The upper boundary of the $\mathcal{R}>0$ region satisfies  $F_2=\frac{1}{12} \left(\frac{d-3}{d-5}\right)$.  For $d=6$, the red line coincides with this line, i.e., $F_2=1/4$ and $\mathcal{R}=0$ (``ideal gas''-like). As $d$ is increased, the $F_2=1/4$ line shifts into the shaded region (more accurately, the region changes while the line stays fixed), so that in the large $d$ limit the red line and black line coincide, as shown in the right most plot for $d=10^4$.}
	\label{pic}
\end{figure*}

The Ruppeiner metric\footnote{Ruppeiner metric is a special case of Fisher information metric. See, e.g., \cite{0604164,0706.0559,1808.08271}. For a brief introduction to its mathematical structures, see \cite{1507.06097}.} \cite{rup} is essentially the Hessian of the entropy function of the black hole hairs,
\begin{equation}
g^R_{ij}:=-\partial_i\partial_j S(M,N^a),
\end{equation}
interpreted as a Riemannian metric (thus endowed with the usual Levi-Civita connection) in the phase space.
Here $M$ is the energy (black hole mass for our case), and $N^a$ are other extensive variables of the system (such as black hole charge and angular momentum). It has been applied to study the thermodynamics of black holes, in particular thermodynamic instability and phase transitions \cite{0802.1326,1309.0901,0304015,0510139,0611119,1001.2220,1004.5550,1507.06097,1008.4482,1005.4832,2009.00291,2010.03310,2107.14523}. The scalar curvature of the Ruppeiner metric, $\mathcal{R}$, plays a crucial role in these analyses. The scalar curvature is positive if the underlying statistical interactions of the thermodynamical system is repulsive, and likewise it is negative for attractive interactions \cite{0802.1326,1007.2160}.
These are referred to as being ``Fermi-like'' and ``Bose-like'', respectively, in\footnote{Note the definition of their Ruppeiner metric -- hence also the curvature scalar -- differs by an overall sign from our convention.} \cite{0808.0241,oshima}.  
Furthermore the larger the value of $\mathcal{R}$, the less stable the system \cite{J1,J2}. A phase transition can be expected when $\mathcal{R}$ diverges\footnote{However, phase transition can also occur when there is no such divergence. See \cite{2103.00935} and the discussion in \cite{brody}.}, while $\mathcal{R}=0$ corresponds to an ``ideal gas''-like behavior.

For the singly rotating Myers-Perry black holes, the Ruppeiner metric is 2-dimensional (its explicit form is not useful for our discussion), and we have \cite{0510139}
\begin{equation}
\mathcal{R} = -\frac{1}{S} \frac{1-12\cdot\frac{d-5}{d-3}\frac{J^2}{S^2}}{\left(1-4\cdot\frac{d-5}{d-3}\frac{J^2}{S^2}\right)\left(1+4\cdot\frac{d-5}{d-3}\frac{J^2}{S^2}\right)}.
\end{equation}
The properties of this scalar curvature and its relation to black hole thermodynamics have been investigated rather thoroughly in the aforementioned literature. 
In particular, as pointed out in \cite{0510139}, in $d=4$ the scalar curvature diverges along the curve $J^2/S^2=1/4$ in the $(S,T)$-plane, which corresponds to the extremal limit.

In view of the appearance of $1/4$, we are motivated to define another quantity by $F_2:=J^2/S^2$ (and refer to the Hookean law by $F_1$ when there is a risk of confusion) and explore its behavior in higher dimensions. Again, we note that $d=5$ is special because $\mathcal{R}$ simplifies to $-1/S$, which becomes divergent in the extremal limit.
Note that $\mathcal{R}<0$ for both $d=4$ and $d=5$. However, for $d \geqslant 6$ the curvature diverges not in the extremal limit (which does not exist), but rather along the curve  
\begin{equation}
F_2 := \frac{J^2}{S^2} = \frac{1}{4} \left(\frac{d-3}{d-5}\right).
\end{equation}

It is even more illuminating if we plot the region that has positive $\mathcal{R}$ in the $(S,T)$-plane, as shown in Fig.(\ref{pic}). The line $F_2=1/4$ that would correspond to the maximum force (if indeed the conjecture was correct and applicable to $F_2$) marks the boundary of the positive $\mathcal{R}$ region in $d=6$, as opposed to marking the extremal limit in $d=4$. As dimensionality increases, the line shifts into the shaded region (or rather the line stays fixed but the shaded region changes). In the large $d$ limit \cite{1302.6382,1504.06613,2003.11394}, it coincides with the phase transition line that marks the lower boundary of the positive $\mathcal{R}$ region. Thus we see that the maximum force bound $F=1/4$ is relevant for black hole thermodynamics. 

\section{Discussion: The Nature of the Hookean Law}

In the context of singly rotating Myers-Perry black holes, the Hookean law ``$F=kx$'' -- which is thermodynamical in nature -- satisfies the maximum force conjecture. In $d=4$ and $d=5$ the maximum value $1/4$ is attained. This is itself surprising, since the Hookean force is not dimensionally a true force. In general, the coefficient $1/4$ for the strong form of the maximum force conjecture does not carry over to other physical quantities. For example -- upon restoring $c$ and $G$ for clarity -- the (conjectured) maximum luminosity is $c^5/G$ \cite{1803.03271} and the maximum Bondi mass loss rate seems to be $0.382 ~c^3/G$ \cite{2109.05973}.

In $d \geqslant 6$, the Hookean force is further bounded by a dimensional dependent value (all smaller than $1/4$). The maximal values of $F$ in $d \geqslant 5$ correspond to Emparan-Myers fragmentation. Black hole spacetimes that are known to fragment are rare in the literature, but our investigation suggests that the relation between the Hookean law and fragmentation is not necessarily true in modified gravity\footnote{This is not surprising since fragmentation is associated with the black hole entropy, which depends on the gravitational field equations (so entropy need not be proportional to horizon area), whereas the Hookean law depends on the Hawking temperature, which is purely kinematical \cite{9712016,0106111}.}, or when matter fields are present. Nevertheless, in the presence of a negative cosmological constant, a 5-dimensional Kerr-AdS black hole that fragments into two asymptotically flat Schwarzschild pieces -- if indeed such a process can happen -- also exhibits the aforementioned surprising property. Generalization to multiply-spinning case is not straightforward due to the constraint on the number of ultra-spinning directions from cosmic censorship (i.e., turning on more angular momenta generically prevents a black hole from becoming super-spinning). That is, not all directions can be ultra-spinning. Hence we would have at least an addition rotation parameter and the analysis becomes complicated. This interesting direction is left for future investigations.
We can, however, already observe that if the Myers-Perry black holes have $n \geqslant 2$ (but equal) angular momenta $J$, the maximal bound for $F_2$ would be different. For example, in $d=6$ and $n=2$, the extremal limit is $J/S=\sqrt{3}/2$, and hence $F_2=3/4 > 1/4$. More generally, if $2n > d-3$, then the extremal limit satisfies
\begin{equation}
F_2^{\text{ext}}=\frac{1}{4}\left[\frac{d-3}{2n - (d - 3)}\right].
\end{equation}
This expression can exceed $1/4$ if $d-3 > n$.

For singly-spinning Myers-Perry black holes, the Ruppeiner thermodynamic geometry approach \cite{rup,1507.06097,0510139,0611119,1001.2220,1004.5550} reveals an interesting property in the phase space $(J,S)$, where $J$ denotes the angular momentum and $S$ the entropy. In dimensions $d\geqslant 6$, the positive Ruppeiner scalar curvature region in the thermodynamic phase space is marked by the upper boundary $J^2/S^2=\frac{1}{12}\left(\frac{d-3}{d-5}\right)$ and the lower boundary $J^2/S^2=\frac{1}{4}\left(\frac{d-3}{d-5}\right)$. Remarkably, the upper and lower boundaries correspond to $J^2/S^2=1/4$ when $d=6$ and $d\to \infty$, respectively. Like $F_1=kx$, the dimensionality of $F_2=J^2/S^2$ is also not a force, but we can again refer to the maximum ``force'' conjecture by multiplying with appropriate factors of $\hbar, c$ and $G$.
Interestingly, the lower boundary at which the curvature diverges corresponds to a black hole that suffers from Gregory-Laflamme instability, which is a dynamical instability but is related to thermodynamical instability \cite{0907.2248,1006.1904} by the extended Gubser and Mitra conjecture \cite{1006.1904,0011127,0104071,1201.0463}. 

\begin{figure}[h!]
\centering
\includegraphics[width=.90\columnwidth]{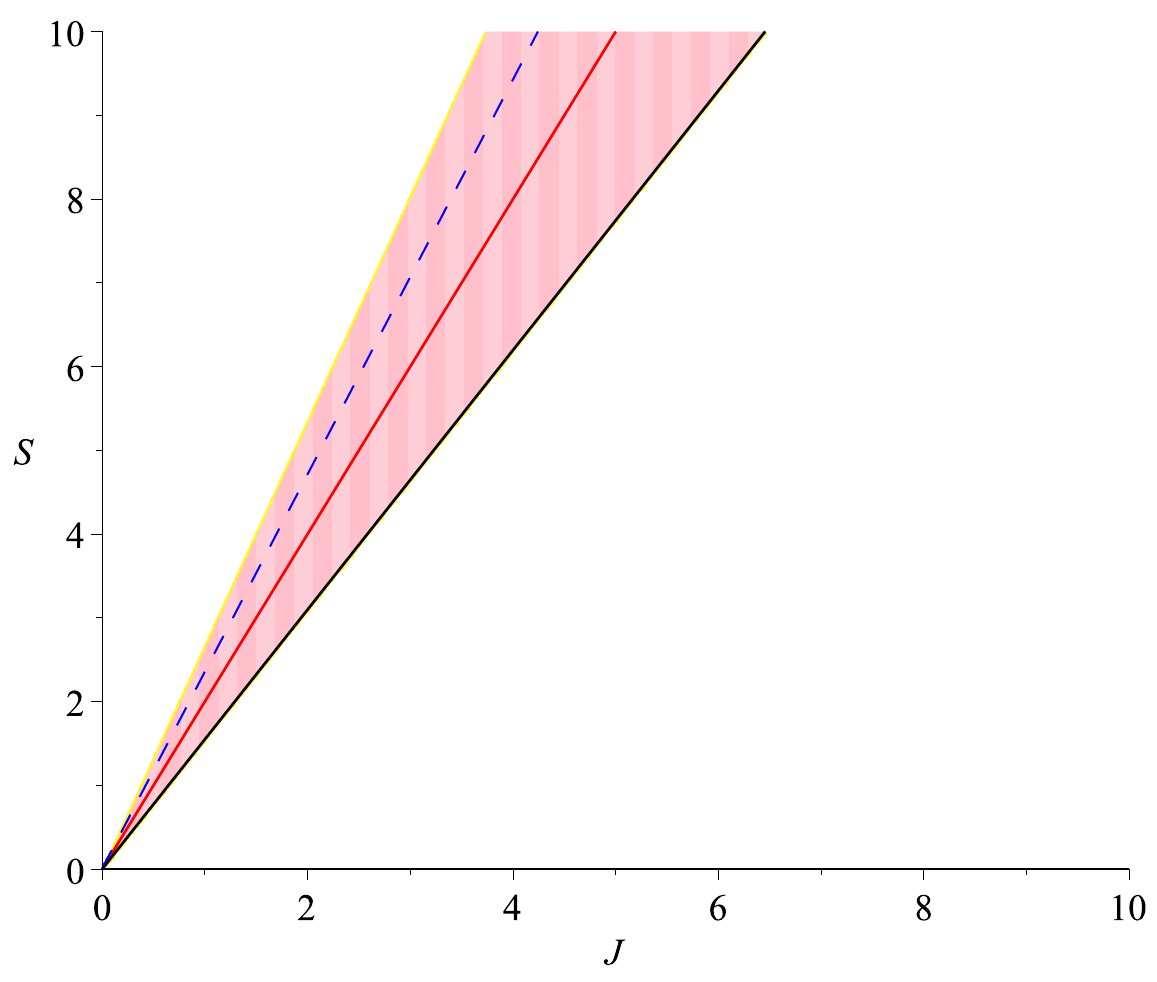} 
\caption{This plot is the same as the middle plot in Fig.(\ref{pic}), which shows the positive Ruppeiner scalar curvature region in pink, and the $F=1/4$ line in red in roughly the middle of the region. 
Here we also show in blue dashed line along which Emparan-Myers fragmentation $F = J^2/S^2 = 0.18 $ occurs.
\label{pic2}}
\end{figure}

It turns out that
the fragmentation line is above the maximum force line $F=1/4$ (See Fig.(\ref{pic2}) for the $d=6$ case). Except in $d=6$ in which the maximum force line corresponds to the upper boundary of $\mathcal{R}>0$ region (and thus the fragmentation line lies in the $\mathcal{R}<0$ region), in all dimensions $d>6$ the fragmentation line also lies in the $\mathcal{R}>0$ region.
From the interpretation of the Ruppeiner scalar curvature, this means that generically the fast-spinning black holes tend to break when the interaction between the underlying microstructures (perhaps ``spacetime atoms'' \cite{1603.08658}) is sufficiently repulsive. Intuitively, if the interaction is attractive, it tends to restore any stretching that occurs to the horizon. As the interaction becomes repulsive, the flattening of the horizon due to large angular momentum eventually breaks the black hole. As the number of dimension increases, the fragmentation line shifts closer towards the divergent $\mathcal{R}$ line.  A possible explanation -- at least intuitively -- is that in higher dimensions, a spacetime atom would have more neighbors that it bonds to in the additional spatial directions, hence  making it more difficult to break the bond\footnote{Interestingly -- though perhaps not directly relevant -- it was previously found that repulsive effect due to the rotation dominates over attraction due to the mass in $d > 5$ at large $r$ \cite{2004.09242}.}. Note that, as mentioned in Sec.(\ref{2}), the numerical values of $F_2$ at which fragmentation occurs \emph{decreases} with dimensions. At first sight this seems to contradict the previous statement that the bonds are more difficult to break in higher dimensions. However,  we cannot really compare  the magnitude of $F_2$ in different dimensions directly as they would have different physical dimensions. What we \emph{can} compare is the various $F_2$ with respect to their respective divergent $\mathcal{R}$-line.
In all cases, fragmentation occurs before the maximum force is attained, so in this sense the maximum force conjecture holds. We remark that Emparan and Myers \cite{0308056} conjectured that even in 5 dimensions, fragmentation could occur at $a^2/\mu \approx 0.72$, which is slightly less than the value the Hookean force attains its maximum $a^2/\mu = 0.75$. This is consistent with the situation in higher dimensions.  

To conclude, our findings suggest that the concept of maximum force appears to hold at least in some thermodynamical contexts of black hole physics. 
In particular, it is remarkable that the Hookean law of black holes -- initially seemingly naively introduced in \cite{1412.5432} -- is related to black hole fragmentation, and thus with the second law of black hole thermodynamics.
However, the maximum force is not necessarily related to the cosmic censorship conjecture, in the sense that the latter might not be applicable (no extremal limit exists) yet the former exists.

Finally, let us remark that in \cite{2005.06809}, Barrow and Gibbons argued that there is \emph{no} maximum force in dimensions above 4. 
Nevertheless, we see some significance of the maximum value $1/4$ in higher dimensions in the examples discussed in this work. We propose a possible explanation for this discrepancy: perhaps the maximum force conjecture only applies to some classes of forces (and by multiplying with appropriate factors of $G,c,\hbar$ -- other physical quantities) in the context of black hole thermodynamics\footnote{Incidentally, we have defined the spring constant in $F=kx$ for higher dimensions using the temperature expression in Eq.(\ref{temp}). This $k$ does \emph{not} satisfy $k=m\Omega_+^2$ in higher dimensions, despite this equation inspired the term ``spring constant'' in 4-dimensions. 
It is also interesting to note that in \cite{0210109}, Gibbons already noticed that in 4-dimensions, the extremal limit of Kerr black holes can be recast into the maximum force form, but remarked that this would fail in higher dimensions due to -- in his words -- ``the different dependence of the gravitational force on separation''. We now see that what \emph{does} generalize is the Hookean law, though the spring constant lost the nice $m\Omega_+^2$ form in higher dimensions.}. One possibility is that it has to do with the forces between ``spacetime atoms'' or ``microstructures'' that underlie black hole horizon, whatever that may be. The fact that the same maximum force conjecture happens to hold in the other contexts explored in 4-dimensions, including the actual gravitational force between two masses studied in \cite{2005.06809}, may be a coincidence. If it is not a coincidence, perhaps this is a hint that gravity in 4 dimensions is in some sense more ``thermodynamical'' than gravity in higher dimensions. If so, this could be relevant to the emergent gravity program \cite{1110.0686,1207.2504,1410.6285,1711.10503}, and perhaps even to the longstanding question on why our Universe is 4-dimensional. 

The fact that the result is not universal, i.e., that fragmentation can occur at values beyond $1/4$ for other black hole solutions (e.g., non-asymptotically flat, modified gravity, or with matter fields) should not be viewed as a shortcoming; it may instead provide a new window to probe these microstructures if we take the spring analogy seriously. Usually a spring can be extended beyond the Hookean limit and still be elastic. It becomes permanently deformed once the displacement is too big and eventually breaks (fragmentation). We speculate that perhaps $F=1/4$ marks the end of the Hookean (linear) regime. In our examples, fragmentation happens while the black hole is still in this regime. The shape of the force as a function of displacement beyond the Hookean regime depends on the material. Perhaps different black hole solutions would allow us to probe how different modifications to the black hole geometry affect the underlying microstructures, beyond the linear regime. This may shed some light on the nature of black hole microstructures, which in turn would help us to understand black hole entropy better. After all, even with all the different approaches from the quantum gravity community to derive the Bekenstein-Hawking entropy, some of its properties remain rather obscure; for example, why would adding angular momentum or electric charge \emph{decrease} the entropy? Why would the simplest solution with most symmetry (the Schwarzschild black hole) maximize the entropy?

A side remark concerning Ruppeiner geometry is in order: Ruppeiner's interpretation of the sign of the scalar curvature $\mathcal{R}$ as being ``Fermi-like'' or ``Bose-like'' interactions was based on how the volume of a small ball grows with radius \cite{1007.2160}, which is more natural in the statistical mechanics models he was investigating. For the black hole case, however, the Ruppeiner geometry is Lorentzian, so in place of small balls we would have small causal diamonds \cite{1507.06097}. Although the same sort of interpretations have been made and widely used in the literature, it is still unclear how to make precise sense of this. Would a Wick-rotation in the phase space help?

It is interesting to note that, historically, Robert Hooke had wanted to describe the behavior of not just actual springs, but also ``springy bodies'', which include both solid and fluid \cite{hooke}. Perhaps it would not surprise him that black holes -- which admit hydrodynamical descriptions \cite{0712.2456,0905.4352,1101.2451,1107.5780} -- exhibit properties that can be described by his law.  \newline

\begin{acknowledgments}
Thanks in memory to the late John Barrow.  MG acknowledges funding from state-targeted program ``Center of Excellence for Fundamental and Applied Physics'' (BR05236454) by the Ministry of Education and Science of the Republic of Kazakhstan, and the
FY2021-SGP-1-STMM Faculty Development Competitive Research Grant No.021220FD3951 at Nazarbayev
University. This work is supported in part by the Energetic Cosmos Laboratory.
YCO thanks the National Natural Science Foundation of China (No.11922508) for funding support. 
\end{acknowledgments}

{}

\end{document}